\documentclass[prb,preprint]{revtex4}
 \usepackage{epsfig}

\usepackage{dcolumn}
\usepackage{bm}

\begin{document}


\title{ 
Combustion Process  in a Spark Ignition 
Engine: Dynamics and Noise Level Estimation}

\author{T. Kami\'nski and M. Wendeker}
\affiliation{Department of Combustion Engines,  Technical University
of
Lublin, \\
Nadbystrzycka 36, PL-20-618 Lublin, Poland}

\author{K. Urbanowicz}
\affiliation{Faculty of Physics, Warsaw University of Technology, \\
Koszykowa 75, PL-00-662, Warsaw, Poland.}

\author{G. Litak}
\affiliation{ Department of Mechanics, Technical University of Lublin,\\
Nadbystrzycka 36, 20-618 Lublin, Poland}
\date{September 30, 2003}

\begin{abstract}
We analyse the experimental  time series of 
internal pressure
in a four cylinder spark ignition
engine.
In our experiment, performed for different spark advance 
angles, apart from usual cyclic changes of engine pressure 
we observed oscillations. These oscillations are with longer time scales 
ranging from one to several hundred engine cycles depending on
engine working conditions. 
Basing on the pressure time dependence we have calculated the heat 
released per 
cycle. Using the time series of heat release to calculate 
the correlation coarse-grained
entropy we estimated the 
noise 
level for internal combustion process.   
Our results show that for a smaller spark advance angle  the system is more
deterministic. 
\end{abstract}

\pacs{PACS numbers:
             05.40.C,  
             05.45.T   
             82.40.B   
}           

\maketitle

\noindent {\bf 
A combustion process in spark ignition engines
is known as  nonlinear and noisy one.
Combustion
instabilities are occurring
as a cycle-to-cycle variations of internal cylinder pressure
effecting directly on the power output.
Examination of these variations can lead
to better understanding of their sources and help in their
eliminations in future.
Improving engine efficiency requires achieving better  combustion conditions without introducing  additional disturbances.  
In the present paper we analyse the dynamics and
estimate the noise level in combustion process
basing on experimental time series of internal pressure
and calculated from them heat release. In the following analysis
we apply the nonlinear multidimensional methods which can distinguish
random variations from a deterministic behaviour. 
}

\section{Introduction}

\begin{figure}
\vspace*{0.0cm}

\centerline{\epsfig{file=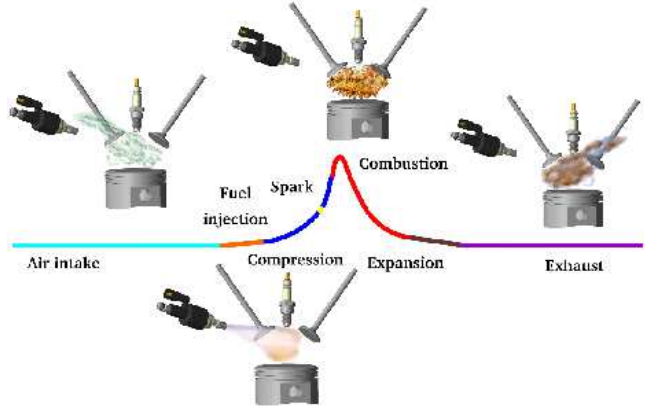,width=8.5cm,angle=0}}
\caption{\label{fig_one}
Schematic picture of  a combustion cycle in a four stroke spark ignition 
engine. 
}
\vspace{-1cm}
~
\end{figure}

Combustion in four stroke spark ignition (SI) engines is a complex 
cyclic process consisted of air intake, fuel injection, compression, 
combustion, 
expansion and finally gas exhaust phases (Fig. \ref{fig_one}) where burned fuel 
power is transmitted through the piston to the crankshaft. 
In early beginning of SI engine development 
there were observed instabilities of  combustion
\cite{Cle86}. These instabilities are causing fluctuations of the  power 
output 
 making it 
difficult to control \cite{Rob97,Wen99}.
The problems of their sources identification  and 
their  
elimination
have became the main issues in SI engines technologies engineering
and they have not been  
solved up to present time \cite{Hu96}. Among the the main factors of 
instabilities 
classified by Heywood \cite{Hey88} are aerodynamics in the cylinder
during combustion, amounts of fuel, air and recycled exhaust gas supplied 
to  the cylinder and a local mixture composition near the spark plug.

Recently, Daw {\em et al.} \cite{Daw96,Daw98} and Wendeker {\it et al.}
\cite{Wen03}
have done  the nonlinear analysis of such process.
Changing an advance spark angle  they observed  the
considerable increase of pressure fluctuations level
\cite{Wen03} claiming that it is due to nonlinear
dynamics of the process.  In the other work \cite{Wen04}
Wendeker and coworkers have
proposed intermittency mechanism to explain
the rout to eventually chaotic combustion.

Prompted by these findings we decided to analyse the
the correlation entropy of the combustion process in different working 
conditions
of the engine. 
With a help of entropy 
produced by the dynamical system 
we can quantify the level of measurement or  dynamical noise 
\cite{Urb03}.
In the present paper we shall start our analysis from  examining
experimental pressure time series. 

It should be noted that pressure is the best known quantity to analyse 
engine dynamics.
Cylinder pressure together with volume data can be used to
obtain indicated mean effective
 pressure (IMEP),  calculate the engine torque, indicated efficiency and 
also  burn
rate, bulk temperature and heat release.
Moreover, statistical analysis of the pressure data can also
 provide information about combustion process stability.

However, in practice, It is not easy to perform direct measurement of  
pressure
\cite{pressure}, as one needs a good sensor persistent to hight 
temperatures to be placed inside the engine 
cylinder.
Therefore to obtain information about pressure some researchers developed 
alternative non-direct measurements procedures
\cite{Ant02}. 

In our case we have been dealing with  novel pressure
fibre optical sensors \cite{sensor}. 
Due to
applying  them 
noise from measurement is very low, comparing to traditional
piezo-electric ones \cite{Lit03}. 
This enabled us to
examine the dynamics more effectively than it was possible in earlier 
investigations.

The present paper is divided into 6 sections. After present introduction 
(Sec. 1),
we will provide the
description of our experimental standing and measurement procedure in Sec. 
2.
There we present some examples of  cycle-to-cycle variations in  
pressure inside one of cylinder.
In Sec 3. we examine the pressure with more detail. 
We also perform   spatio-temporal 
analysis 
comparing 
the fluctuations of 
pressure in succeeding cycles for  different advance angles. In Sec. 4 we 
calculate the heat release per cycle. 
Finally in Sec. 5 we analyse  its time dependence  
and  show our main result i.e. level  of noise. We end up with conclusions 
and last remarks (Sec. 6).

\section{Experimental Facilities and Measurements of Internal Pressure}

\begin{figure}
\centerline{\epsfig{file=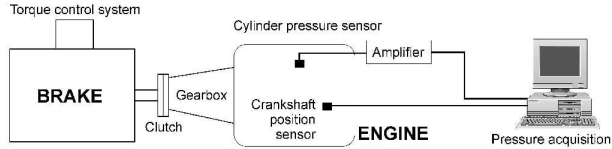,width=8.2cm,angle=0}}  
 \caption{\label{fig_two}
Experimental stand.
}
 \end{figure}

In our experimental stand (Fig. \ref{fig_two}) pressure was measured directly 
inside 
cylinder by the use of the optical fibre  sensor \cite{sensor}.
Such equipment provides one of the most direct measures of combustion 
quality in an internal 
combustion engine.  
Internal pressure data were obtained from Engine 
Laboratory of Technical University of Lublin, where we conducted series of 
tests. 

The 
pressure traces were generated on a
 1998 cm$^3$ Holden 2.0 MPFI engine at 1000 RPM. The data was captured by use of  
NuDAC--TK v.2.0 data acquisition 
and data 
processing 
program \cite{Kam03}.
 The original files contained cylinder pressure at crank  angles 0-720 
degrees. Each of three large file (about 990 MB 
per 
each) contained above 10000 combustion
 cycles. Data was taken  at different spark timings (spark advance angles): 5,15,30 
degrees before top dead center (BTDC). The engine 
speed, 
air/fuel ratio, and throttle
 setting were all held constant throughout  the data collection period. 
Intake air pressure, in inlet pipe, was value 40 kPa. Torque for each of three
 spark timings were adequately: 21, 28 and 30 Nm.

\begin{figure}
\centerline{\epsfig{file=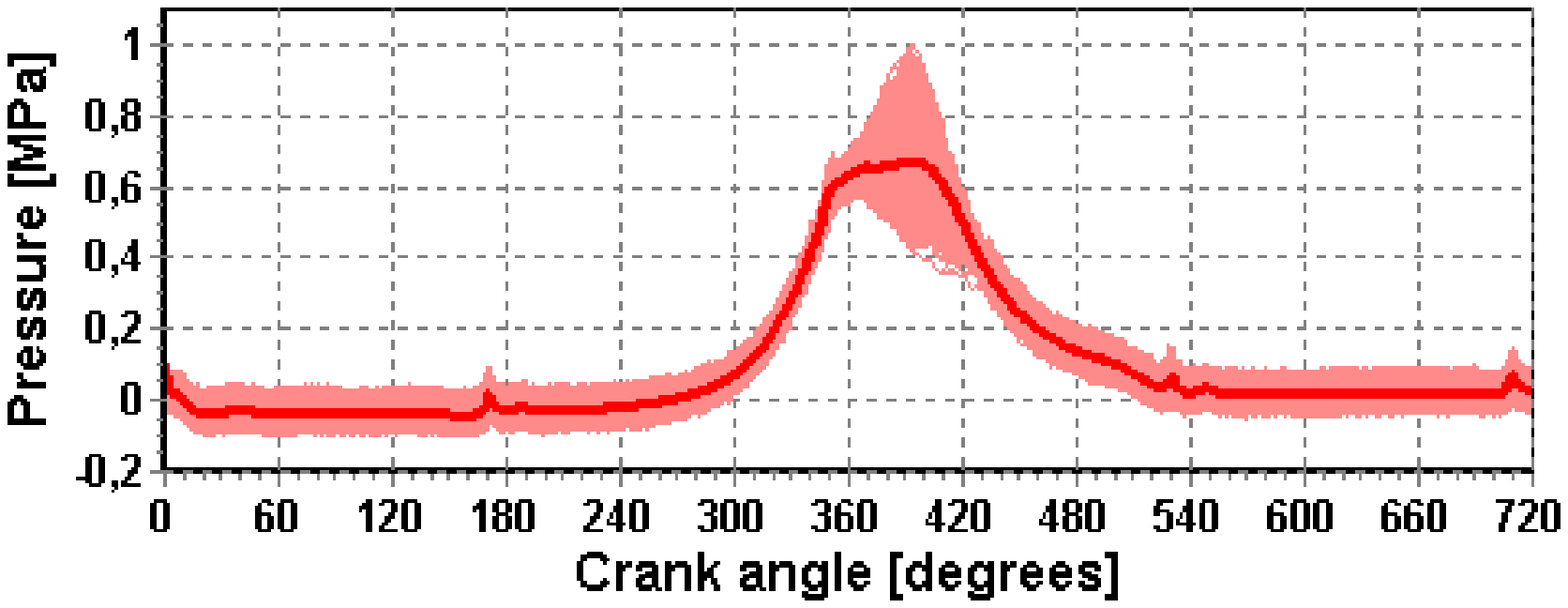,width=9.0cm,angle=0}}

\centerline{\epsfig{file=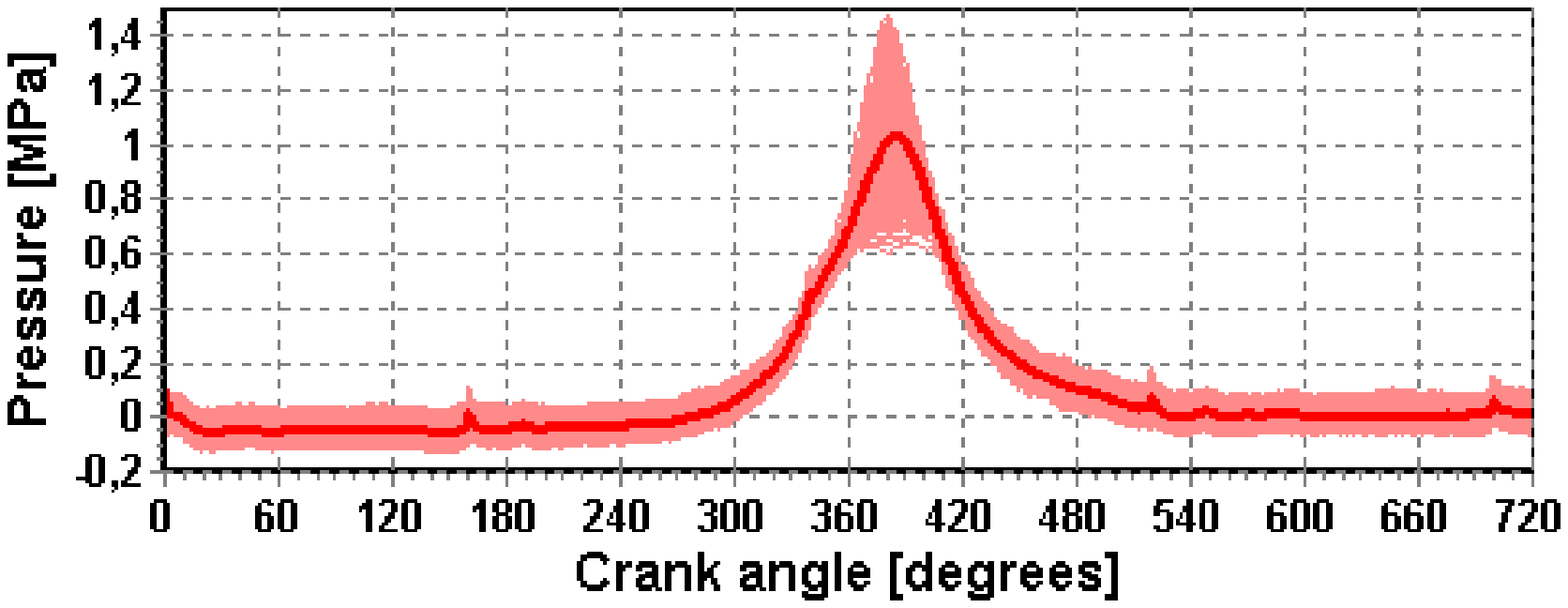,width=9.0cm,angle=0}}

\centerline{\epsfig{file=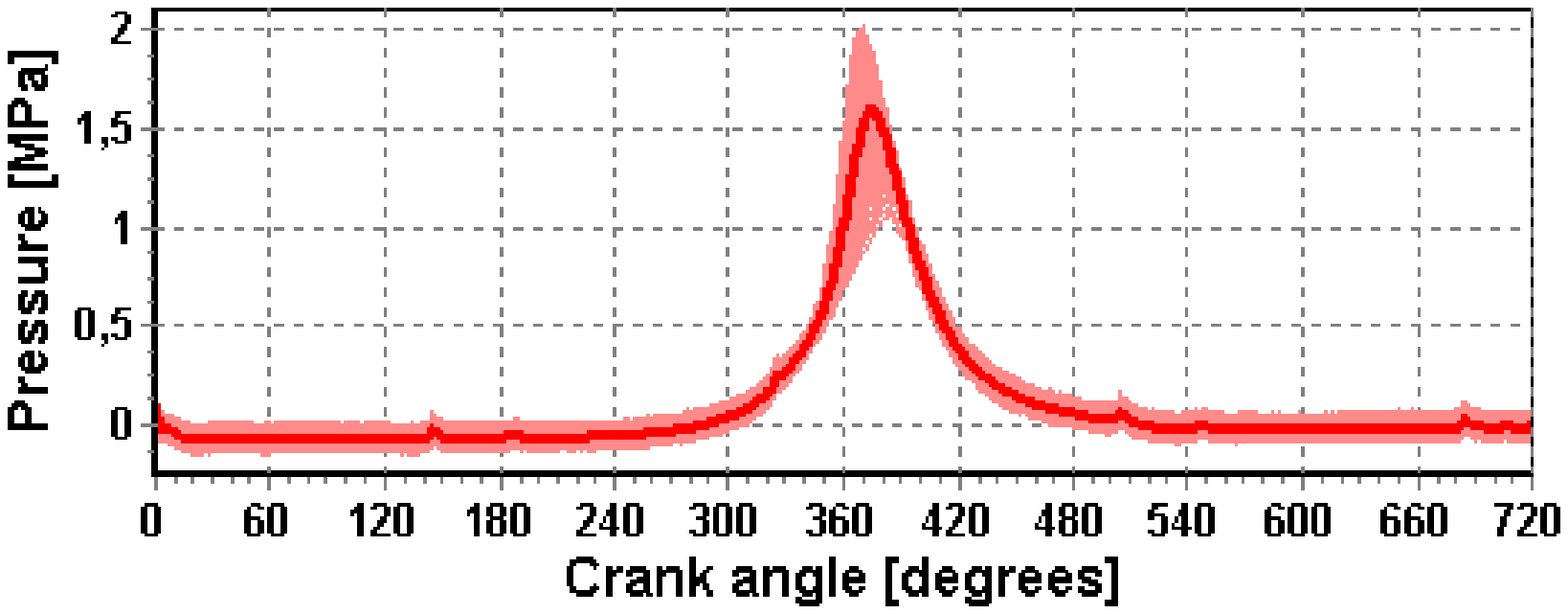,width=9.0cm,angle=0}}

\vspace{-8.5cm} \hspace{-8cm} (a)

\vspace{3.6cm} \hspace{-8cm} (b)

\vspace{3.6cm} \hspace{-8cm} (c)
\vspace{0.5cm}

 \caption{\label{fig_three}
Internal pressure of 1000 combustion cycles against a crank angle for a 
spark advance angle 
$\Delta \alpha_z=5$, 15 and 30 
degrees for Figs. 3a,b and c, respectively. Full lines correspond to the 
average angular pressure.  
}
 \end{figure}

\begin{table}

\caption{\label{tableone} 
Definitions of variables and symbols used in the
paper.
}
\vspace{1cm}
\hspace{2cm}
\begin{tabular}{c|c}
 \hline
position of the piston & $h$ \\
internal cylinder pressure & $P$ \\
actual cylinder volume & $V(\alpha)$ \\
Heaviside step function
 &  $\Theta(z)$ \\
heat released  & $Q$ \\
heat released in particular cycle $i$  & $Q_i$ \\
heat released vector in embedding space & ${\bf Q}$ \\
spark advance angle \\ $\Delta \alpha_z$
embedding time delay in cycles&  $m$ \\
cycle number & $i,j$ \\
embedding dimension & n \\
number of considered points in time series & $N$ \\
loading torque & $F$ \\
crank angle & $\alpha \in [0,720^o]$ \\
threshold & $\varepsilon$ \\
correlation integral & $C^n$ \\
coarse-grained correlation integral & $C^n (\varepsilon)$ \\
correlation entropy & $K_2$ \\
coarse-grained correlation entropy & $K_2(\varepsilon)$ \\
calculated from time series \\coarse-grained entropy & $K_{noisy}$ \\
correlation dimension & $D_2$ \\
Noise-to-Signal ratio & NTS \\
standard deviation of data & $\sigma_{DATA}$\\
error function & Erf$(z)$ \\
fitting parameters & $\chi$ , $a$, $b$ \\
standard deviation of noise & $\sigma$ \\
 \hline
\end{tabular}
\end{table}
\begin{table}
TABLE I: continuation

\vspace{1cm}
\hspace{2cm}
\begin{tabular}{c|c}
 \hline
cylinder diameter & $D= 86$ mm \\
crank  radius    & $r= 43$ mm \\
connecting-rod length  & $l= 143$ mm \\
heating value of the fuel  & $W_u=43000$ kJ/kg \\
compression ratio & $\varepsilon=8.8$ \\
Poisson constant  & $\kappa=\frac{c_p}{c_v}\approx 1.4$ \\
mass burned in cycle $i$ & $M_i$\\
autocorrelation function & $AC(j)$ \\
output torque & $S$ \\
\hline

\end{tabular}

\end{table}

\begin{figure}
\centerline{\epsfig{file=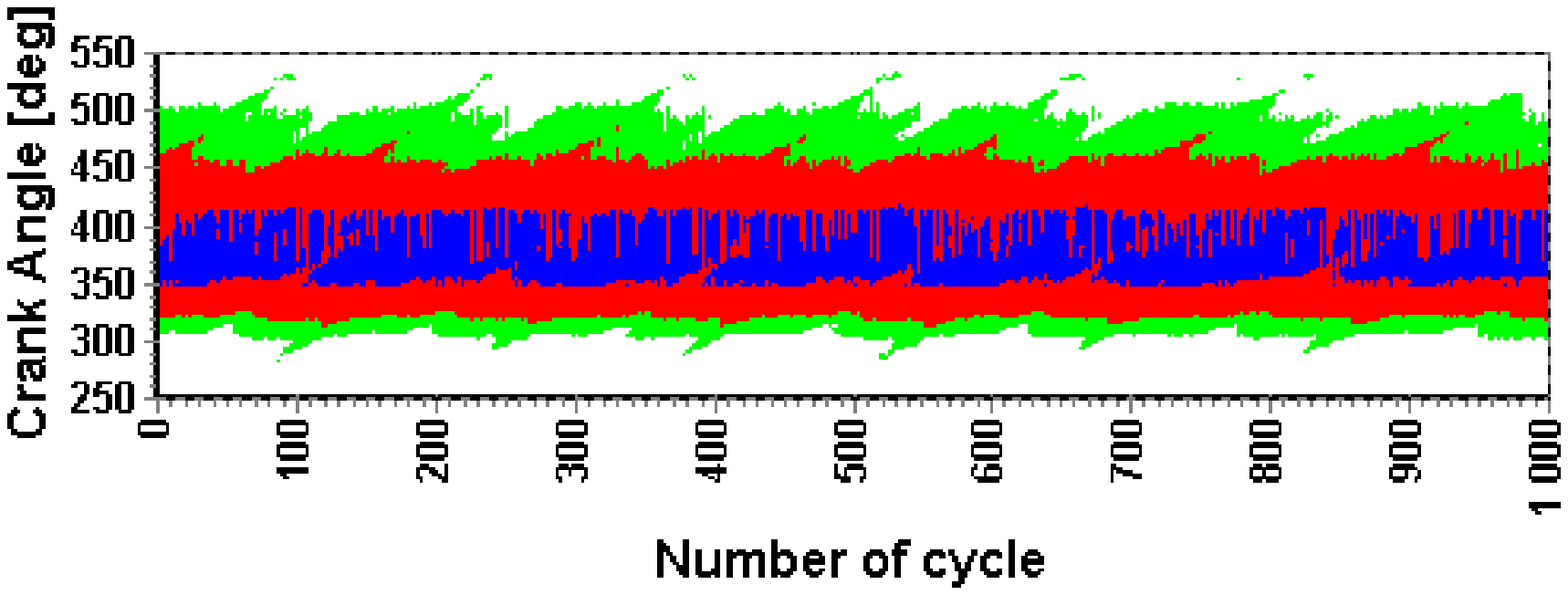,width=9.0cm,angle=0}} 

\centerline{\epsfig{file=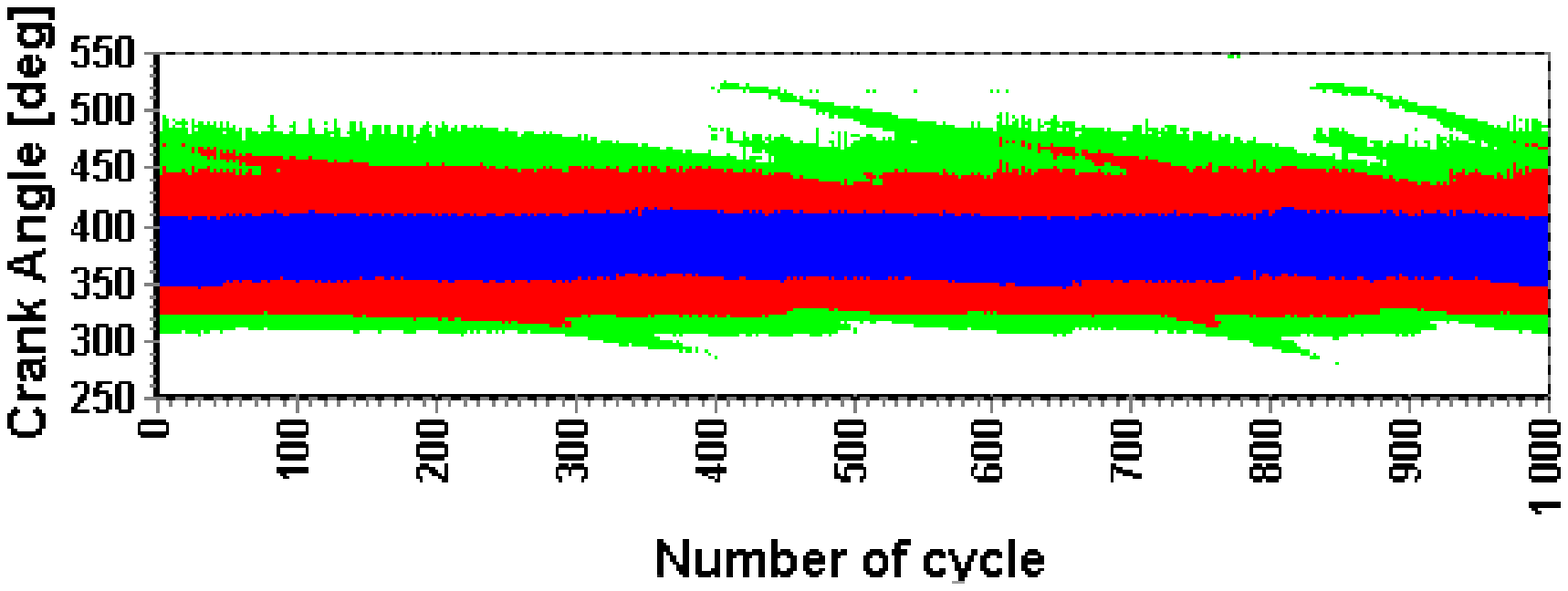,width=9.0cm,angle=0}}

\centerline{\epsfig{file=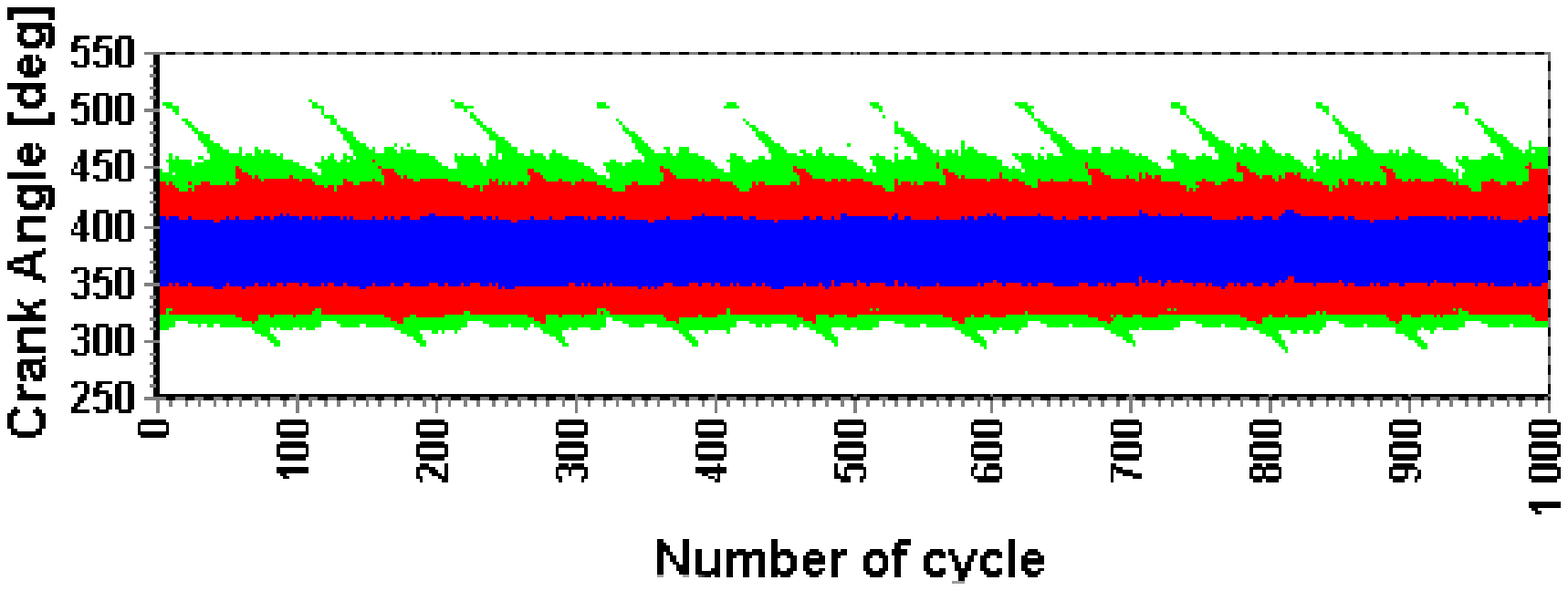,width=9.0cm,angle=0}}

\vspace{-8.3cm} \hspace{-8cm} (a)

\vspace{3.4cm} \hspace{-8cm} (b)

\vspace{3.4cm} \hspace{-8cm} (c)
\vspace{0.5cm}

 \caption{\label{fig_four}
Four colour Spatio-temporal corresponding 
combustion process parameters as in Fig. 3a-c ,respectively. Each colour
correspond to one of four interval of $\left[P_{min},P_{max}\right]$: 
white 
$\left[P_{min},P_1\right]$, 
green 
$\left[P_1,P_2\right]$, red  $\left[P_2,P_3\right]$ and blue $\left[P_3, 
P_{max}\right]$ ($P_{min}=-0.2$MPa, $P_1=0.1$MPa, $P_2=0.2$MPa, 
$P_3=0.6$MPa, $P_{max}=2.0$MPa).
 }
 \end{figure}

To perform signal analysis we needed large enough data. In this aim 
we measured 10000 cycles for each of three spark advance angles  
$\Delta \alpha_z$. The results for first 1000 cycles are shown in Figs. 
3a-c. Note that depending om an advance angle we have more or less 
broadened region of pressure fluctuations. The full line shows the 
pressure averaged over the first 1000 cycles. It is increasing with 
growing $\Delta \alpha_z$ and reaches its highest value 
for $\Delta \alpha_z=30^o$. 

Note also , in our for four stroke engine the combustion period (Figs.
1,3)
corresponds exactly to the double period of
the crankshaft revolution synchronised with a single spark ignition.
Every combustion cycle starts  with  initial conditions given by a 
mixture of air and fuel.
All of succeeding cycles are separated by gas exhaust and intake stroke 
phases. That process is in general nonlinear and can be also mediated by 
stochastic disturbances coming from e.g. non homogenous spatial 
distribution of fuel/air ratio. After combustion exhaust gases are mixed 
with fresh portions of fuel and air.

Therefore the residual cylinder gases after each combustion cycle 
influence the
process in a succeeding cycle leading to different initial conditions
of air, fuel and residual gas mixture contents.

\section{Analysis of Pressure}

During the combustion process the internal volume of engine cylinder
is driven kinematically by the piston.
In a consequence of above the internal it
changes also periodically as a function of crank angle $\alpha$ and  
satisfying the relation
\begin{equation}
V(\alpha) = \pi \frac{D^2}{4}h +\pi \frac{D^2}{4}2r\frac{1}{\varepsilon-1},
\label{eq1}
\end{equation}
where the piston position $h$
\begin{equation}
h = r(1-\cos \alpha) +l \left(1- \frac{r}{l} \sqrt{\frac{l^2}{r^2} -  \sin^2 
\alpha} \right),
\label{eq2}
\end{equation}
and constants  $r$, $l$, $D$ as well as $\varepsilon$  are
defined in Tab. \ref{tableone}

In some sense the combustion initiated by ignition in each engine cycle is an
independent combustion event.
Such
events are separated by the processes of exhaust and
 intake dependent on the
 amount of  combusting fuel mass and quality of newly
prepared  fuel-air mixture.
To  illustrate this effect we are showing in Figs 4a-c  spatio-temporal 
plots corresponding to first 1000 cycles of our pressure time series 
for different advance angle $\Delta \alpha_z=5$, 15, 30 degrees, 
respectively. Each colour in Figs. 4a-c
correspond to one of four interval of $\left[P_{min},P_{max}\right]$:
white
$\left[P_{min},P_1\right]$,  
green
$\left[P_1,P_2\right]$, red  $\left[P_2,P_3\right]$ and blue $\left[P_3,
P_{max}\right]$.
One can easily see that the pressure signal, especially in Fig. 4c seems to 
change in some regular manner of a time scale of about 100 cycles.
Similar feature is also visible in Fig. 4a while it is difficult to 
find such regularity in Fig. 4b. However after more careful examination  one can 
identify such a time scale consisting 
of about 450 cycles. 
  
The broad angular interval of fluctuations visible in Fig. 3b.
has its consequences in irregularly border between  
red and blue colours (Fig. 4b).

\begin{figure}
\hspace{1.5cm}

\centerline{\epsfig{file=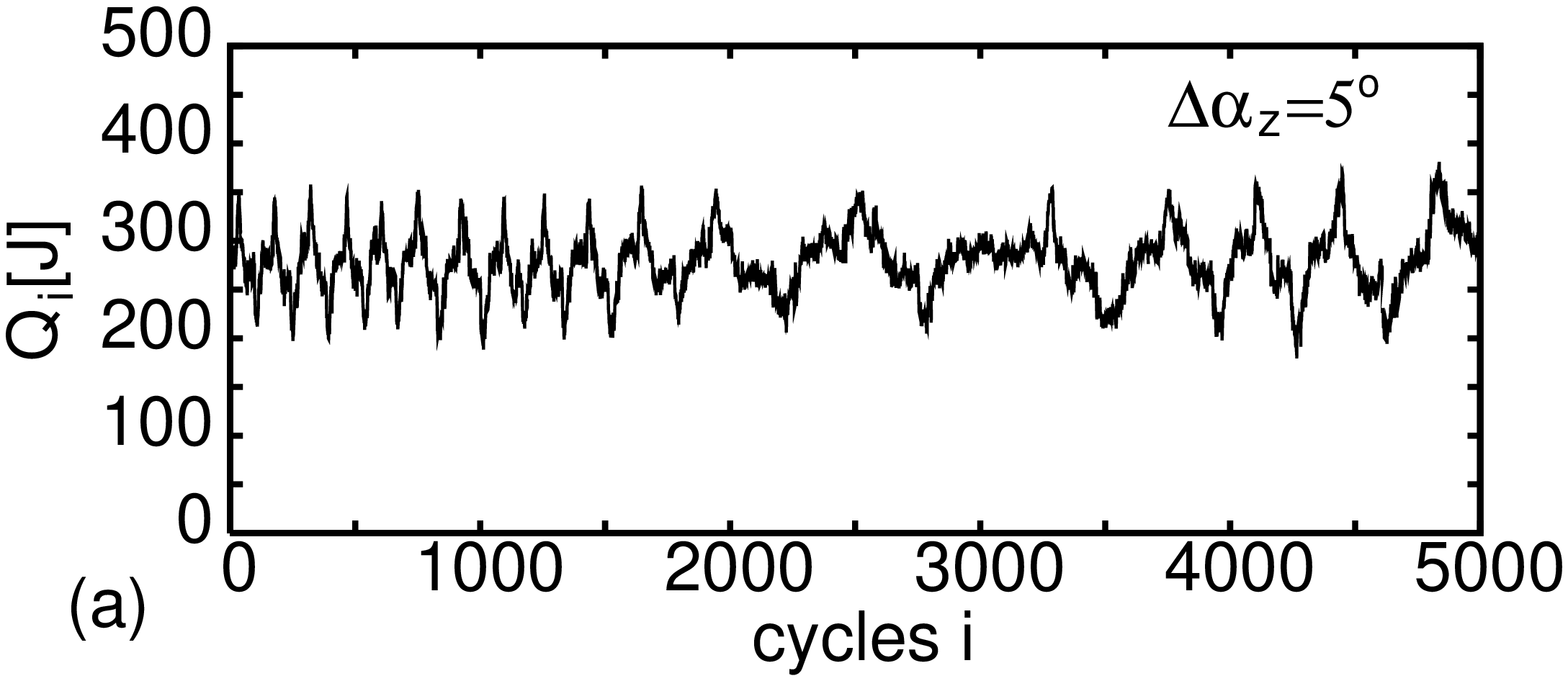,width=4.0cm,angle=-90}}

\hspace{3.5cm}
\centerline{\epsfig{file=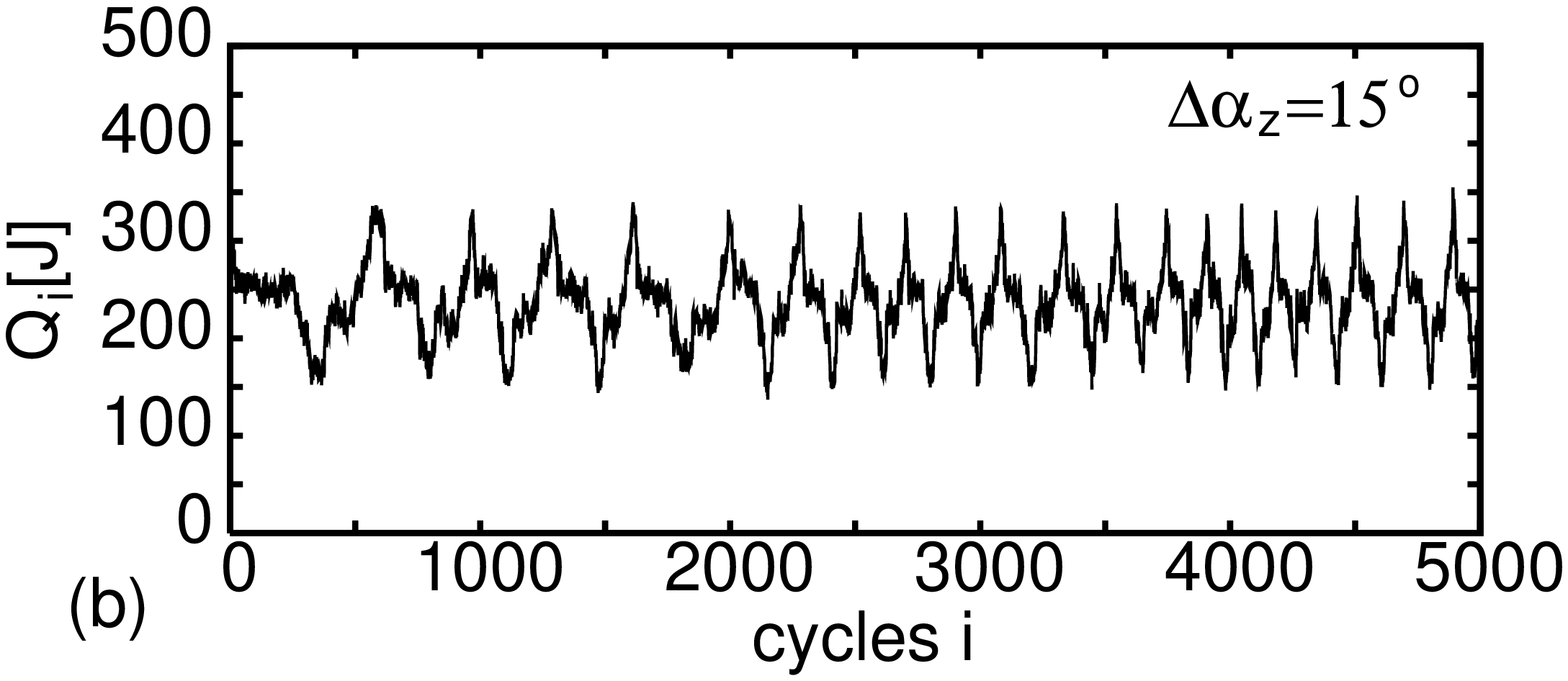,width=4.0cm,angle=-90}}

\hspace{1.5cm}
\centerline{\epsfig{file=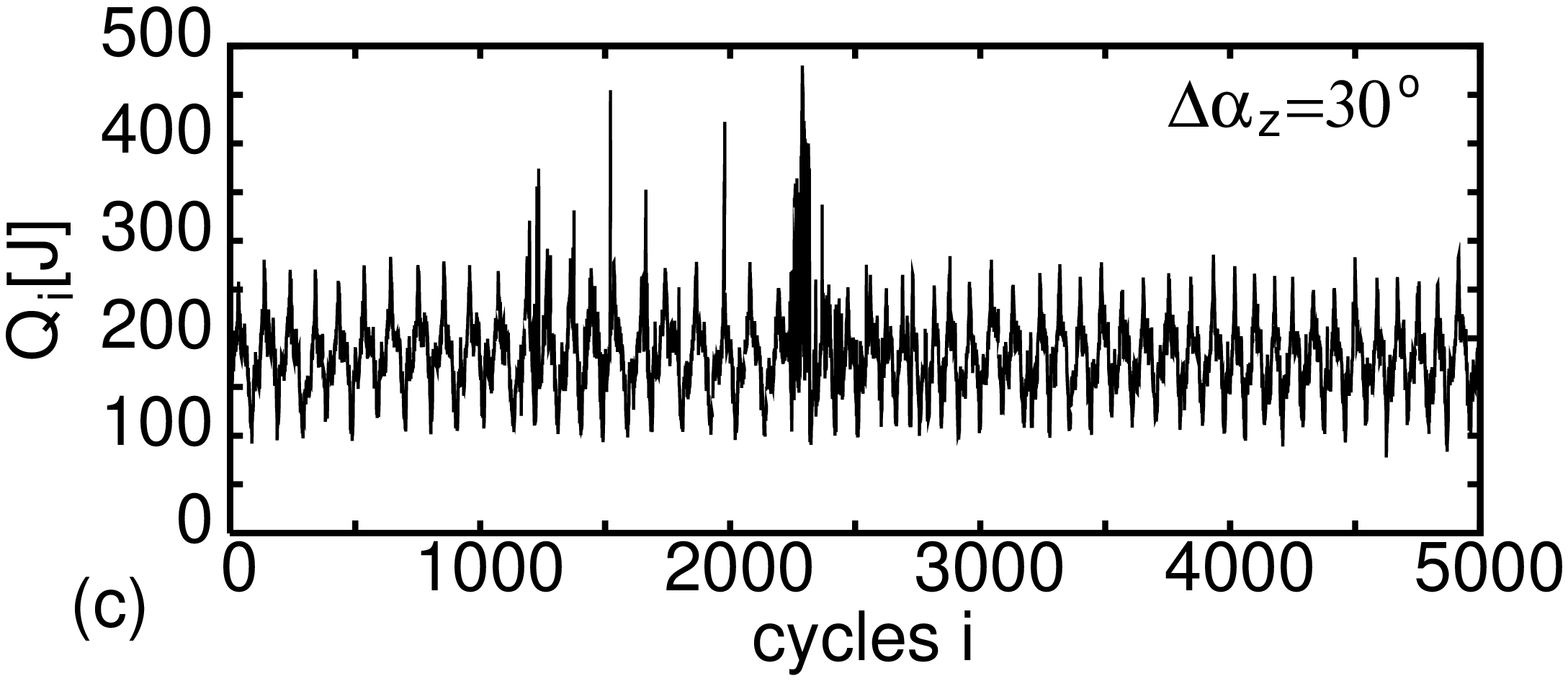,width=4.0cm,angle=-90}}

 \caption{\label{fig_five}
Heat release per cycle versus sequential cycles. 
}
 \end{figure}

\section{variations of Heat release}

To capture the cycle-to-cycle changes in combustion process we decided to 
calculate of heat release in a sequence of combustion cycles.
This quantity is more convenient to examine stability of combustion 
process because enables us to
concentrate on it. In contrast to that internal pressure was is
effected by 
combustion and cyclic compression. Heat release has also practical meaning 
as it is proportional to burned fuel mass.

For an adiabatic process the  heat released from chemical reactions
 during combustion in the engine is 
given 
by an differential
equation coming from the first law of thermodynamics with respect to a crank angle 
$\alpha$: 
\begin{equation}
\frac{{\rm d} Q}{{\rm d} \alpha} =
\frac{\kappa}{\kappa -1} p \frac{{\rm d} V}{{\rm d} \alpha} + 
\frac{1}{\kappa -1}V \frac{{\rm d} 
p}{{\rm d}
\alpha}.
\label{eq5}
\end{equation}

Using it together with the pressure time series and parametric change of 
cylinder volume $V$ Eqs. 1-2
we have done calculation
of heat released in succeeding
cycles $Q_i$.

It is closely related to burned fuel mass in one cycle
\begin{equation}
M_i = Q_i/W_u,
\label{eq6}
\end{equation}
where $W_u$ is heating value of the fuel, listed in Tab. 1.
I should be underlined that the last equation (Eq. 4) neglects the effects of heat exchange 
between the cylinder chamber and its walls. This is in spirit of an adiabatic process assumption
(Eq. 3).  
Of course the consumed mass will be larger because it also depends on the
quality of mixture and combustion process. 

The calculated heat release $Q_i$, for first 5000 cycles is plotted against 
cycles $i$ in Fig. 5a-c for $\Delta \alpha_z=5,15,30$ degrees, 
respectively.

Note that in all cases there is some modulation ranges from one
to few hundred cycles. Interestingly, for a small advance angles 
 $\delta \alpha_z=5^o$ or $15^o$ this modulation evolute 
indicating that the 
system can have quasiperiodic or chaotic nature.
Note also for the first 1000 variations for $\Delta \alpha_z=5^o$ (Fig. 5a)
resembles those for $\Delta \alpha_z=30^o$ (Fig. 5c) while  for
$\Delta \alpha_z=15^o$ the long time scale modulation is different. This is 
consistent with Fig. 4a-c. 
Generally, for $\Delta \alpha_z=30^o$ the oscillations a of higher frequency 
and more 
regular.  
The high values of $Q_i$ in the middle part    
of Fig. 5c ($i \in [1300,2400]$) are connected with measurment 
instabilities.
In that case  however the average value of heat release $<Q_i>$ 
($<Q_i>=270$J - Fig5a
$<Q_i>=237$J - Fig5b $<Q_i>=181$J - Fig5c)
is the 
smallest indicating the lowest burning rate of fuel. In spite of that 
 the output torque, for the same speed of a crankshaft, was relatively 
larger $S=30$Nm (in the case $\Delta \alpha_z=30^o$) comparing to
other levels 21 and 28Nm for  $\Delta \alpha_z=5^o$ and $15^o$, 
respectively.  Obviously, there are better combustion conditions in the last case.

We have also calculated autocorrelation function from the whole 10000 
cycles signal via 
\begin{equation}
AC(j)=  \sum_{i} Q(i)Q(i+j)
\end{equation}
with appropriate normalisation to one. 
The results for all three  advance angles $\Delta \alpha_z$ are depicted in
Fig. 6. One can see that the decay of $AC(j)$ amplitude with growing $j$
is comparable but frequency of modulation is 
different for all these cases. Clearly, this is higher for larger  
$\Delta \alpha_z$.

\section{Estimation of Noise Level from Heat Release Series}

In this section we shall examine the level of noise in heat release time 
series. In this aim we use nonlinear embedding space approach\cite{Kan97}.

In the $n$ dimensional
   embedding space the state is represented by a
vector
\begin{equation}
\label{eq1}
{\bf Q}=
\{Q_i,Q_{i+m},Q_{i+2m},...,Q_{i+(n-1)m} \},
\end{equation}
where $m$ denotes the embedding delay in terms of cycles. The
correlation integral calculated in the embedding space can be
defined as
 \cite{Paw87,Gra83}

\begin{figure} 
\centerline{\epsfig{file=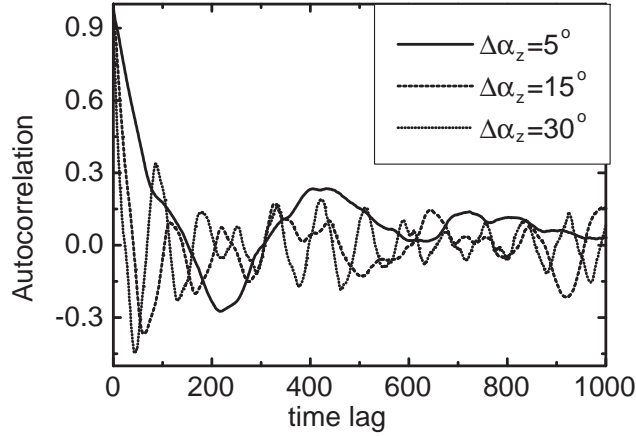,width=9.5cm,angle=0}}
\caption{\label{fig_six}
Correlation function calculated from heat release time series
for different advance angle $\Delta \alpha_z$.}
\end{figure}

\begin{equation}
\label{eq2}
C^n(\varepsilon) = \frac{1}{N^2}  \sum_{i}^{N} \sum_{j \neq i}^{N} \Theta 
( \varepsilon
-|| \mathbf{Q}_{i}-\mathbf{Q}_{j} ||),
\end{equation}
where $N$ is the number of considered points corresponding to
pressure peaks in cycles and $\Theta$ is the Heaviside step
function. For simplicity we use maximum norm. The correlation
integral $C^n(\varepsilon)$ is related to the correlation entropy
$K_2(\varepsilon)$ and the system correlation dimension $D_2$ by
the following formula \cite{Paw87,Gra83}
\begin{equation}
\label{eq4}
 \lim_{n \rightarrow \infty} C^n(\varepsilon)=D_2
\ln \varepsilon -nmK_2(\varepsilon).
\end{equation}  

The coarse-grained correlation entropy can be now be calculated as

\begin{equation}
\label{eq3}
K_2(\varepsilon)=  \lim_{n \rightarrow \infty} \ln \frac{C^n 
(\varepsilon)}{C^{n+1}
(\varepsilon)} \approx - \frac{ {\rm d} \ln C^n (\varepsilon)}{{\rm d}
n}.
\end{equation}
In such a case the correlation entropy is defined in the limit of
a small threshold $\varepsilon$.

In presence of noise described by the standard deviation $\sigma$
of ${\bf Q}_i$ time series,
the observed coarse-grained entropy $K_{noisy}$
\cite{Urb03, Lit03} can be written as
\begin{eqnarray}
\label{eq5}
K_{noisy}(\varepsilon) &=& -\frac{1}{m} {\rm g}\left(\frac{\varepsilon}{2
\sigma} \right)
 \ln \varepsilon + \left[ \chi +b \ln (1-a \varepsilon) \right]
 \nonumber
\\
&\times& \left( 1 +\sqrt{\pi} \frac{\sqrt{ \varepsilon^2/3 +2
\sigma^2} - \varepsilon/\sqrt{3}}{\varepsilon}   \right).
\end{eqnarray}  

\noindent Function g$(z)$, present in the above formula, reads
\begin{equation}
\label{eq6}
{\rm g}(z)=\frac{2}{\pi} \frac{z {\rm e}^{-z^2}}{{\rm Erf}(z)},   
\end{equation}

where Erf$(.)$ is the Error Function. The parameters $\chi$,
$a$, $b$ as well as $\sigma$ are unconstrained. They should be
fitted in Eq. (\ref{eq5}) to mimic the observed noisy entropy
calculated from avaliable data.

\begin{figure}
\centerline{\epsfig{file=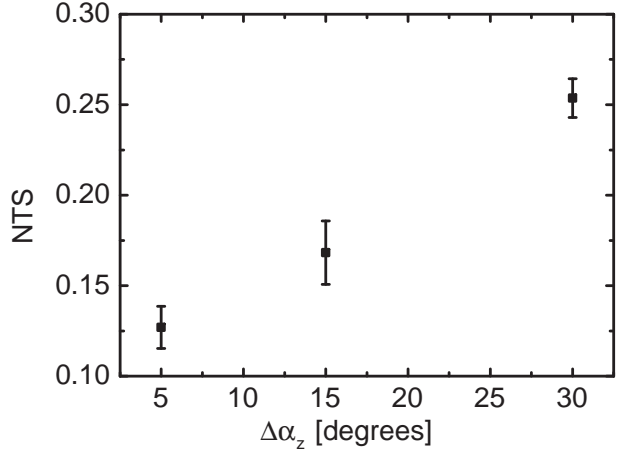,width=9.5cm,angle=0}}
\caption{\label{fig_seven}
Noise to signal ratio NTS  versus a spark advance angle $\Delta 
\alpha_z$.
}
\end{figure}

After application the above method to the heat release times series
we estimated noise calculating  Noise to Signal ratio (NTS):
\begin{equation}
{\rm NTS}=\frac{\sigma}{\sigma_{DATA}},
\end{equation}
where $\sigma_{DATA}$ is the standard deviation of data.

Figure 7 shows NTS versus a spark advance angle $\Delta
\alpha_z$. It appeared that 
for any of examined cases
the noise level is not high.
Starting from a small spark advance angle 
$\Delta\alpha_z=5^o$ the level of noise is the smallest NTS$ \approx 13\%$.
It grows to about 17$\%$ for $\Delta\alpha_z=15^o$ and 
25.5$\%$ for $\Delta\alpha_z=30^o$, respectively. 
In fact easy to note that all points (in Fig. 7) lay on a line.
and ratio may be discribed as increasing linearly with 
 $\Delta\alpha_z$. 

In all considered cases we have to do with a some large scale signal modulation ranging to  
a few hundered engine cycles and
fluctuations of a few tens or more as well as very fast ones (Figs. 4 and 5). 
Using the same method we have examined the natue of very fast fluctuations and
found that they are purely stochastic.
Fluctuation of large time scales may involve in the estimation of noise level to make them 
underestimated by a few percents. This is because the correlated noise can
occur and such a noise is irrelevant to our method. We think that in all three
cases the calculated noise level are biased similary, because of the same
experimental conditions, so the linear increasing behaviour of NTS versus
advance angle $\Delta\alpha_z$
is preserved.  

\section{Remarks and Conclusions} 

In this paper we analysed instabilities of combustion process
We started from analysing pressure time series. Using spatio-temporal 
methods we established that there is a long time scale in fluctuation of 
our experimental 
data. To examine this phenomenon in detail we calculated heat release
and we performed the noise level 
estimation using nonlinear multidimensional methods.
Our results clearly indicate that the noise in the time series 
is the highest for the largest advance angle. In case of 
$\Delta\alpha_z=30^o$ we have got the signal with characteristic 
100 cycles periodicity. 

Heat released in cycle  as a practical parameter  
closely related to burned fuel  mass and enables to follow the stability of combustion
process better. 
Our noise estimation basing on  heat release time series
is more credible than the analysis  pressure histories itself, as 
pressure is effected by volume cyclic compression and expansion phases. 

The method using correlation entropy which we applied here differs from the symbolic  
treatments \cite{Daw00,Daw03} used also for exploration of the engine dynamics by Daw 
{\it et al}
\cite{Daw98}
In their paper the signal was digitised and basing on the
probabilities of 0 1 sequences probabilities the information Shannon entropy 
was estimated.
In our paper we estimate the NTS
ratio by fitting coarse-grained entropy obtained from experimental time series to 
a general formula of correlation entropy evaluated in presence of noise.      
In our case the entropy has its dynamical meaning as a measure uncertainty of 
system state.

One should also note that our present examination 
was limited to only one crank rotational speed 1000 RPM. Using the above 
procedure we now preparing systematic
analysis with the other speeds for a future report.

\section*{Acknowledgements}
Two of authors (KU and GL) would like to thank
Max Planck Institute for 
Physics 
of Complex
Systems in Dresden for hospitality. 
During their stay in Dresden an important
part of data analysis was performed. 
KU has been partially supported by KBN Grant 2P03B03224.


\begin{thebibliography}{99}
\bibitem{Cle86}
D. Clerk, {\it The Gas Engine}, (Longmans, London 1886).

\bibitem{Wen99} M. Wendeker, A.  Niewczas, B. Hawryluk,
{\it SAE Paper} 00P-172 (1999).
  
\bibitem{Rob97}
J.B. Roberts, J.C. Peyton-Jones, K.J. Landsborough,
{\it SAE Paper} 970059 (1997).

\bibitem{Hu96} Z. Hu, 
SAE Paper 961197 (1996).

\bibitem{Hey88}
J.B.  Heywood, {\it Internal combustion engine fundamentals} (McGraw-Hill, New
York 1988).

\bibitem{Daw96} C.S. Daw, C.E.A. Finney, J.B. Green Jr., M.B. Kennel, J.F. Thomas 
and F.T. Connolly,  
{\it SAE Paper} 962086 (1996).

\bibitem{Daw98} C.S. Daw, M.B. Kennel, C.E.A. Finney, F.T. Connolly,
{\it Phys. Rev.} E {\bf 57}, 2811 (1998).

\bibitem{Wen03} M. Wendeker, J. Czarnigowski, G. Litak and K. Szabelski,
{\it Chaos,
Solitons \& Fractals} {\bf 18}, 803 (2003).

\bibitem{Wen04} M. Wendeker, G. Litak, J. Czarnigowski and K. Szabelski,
{\it Int. J. Bifurcation and Chaos} {\bf 14}, (2004).

\bibitem{Urb03} K. Urbanowicz, J.A. Ho\l{}yst, 
{\it Phys. Rev.} E {\bf 67} 046218 (3003).

\bibitem{pressure}
S. Leonhardt, N. Müller, R. Isermann,
{\it IEEE/ASME Transactions on Mechatronics} {\bf 4} 235 (1999).

\bibitem{Ant02} Antoni I, Daniere J, Guillet F. {\it Journal Sound and Vib.}
{\bf 257} 839 (2002).


\bibitem{sensor} M. Wendeker, T. Kami\'nski, M. Krupa, Journal of KONES {\bf 10} 
373 (2003).

\bibitem{Lit03} G. Litak, R. Taccani, R. Radu, K. Urbanowicz, J.A. Ho\l{}yst, M.
Wendeker, A. Giadrossi
{\it Chaos, Solitons \& Fractals}
submitted   (2003).


\bibitem{Kam03} T. Kami\'nski, unpublished.

\bibitem{Kan97} H. Kantz, T. Scheiber, 
(Cambridge University Press, Cambridge 1997).

\bibitem{Paw87} K. Pawelzik and H.G. Schuster, 
{\it Phys. Rev.} A {\bf 35}, 481 (1987).

\bibitem{Gra83}
P.  Grassberger and I. Procaccia,
{\it Phys. Rev. Lett.} {\bf 50}, 346 (1983).

\bibitem{Daw00} C.S. Daw,  C.E.A. Finney and M.B. Kennel
{\it Phys. Rev.} E {\bf 62}, 1912 (2000).

\bibitem{Daw03} C.S. Daw, C.E.A. Finney and E.R. Tracy,
{\it Rev. of Scien. Instruments} {\bf 74}, 915 (2003).



\end{thebibliography}
\end{document}